# Cosmic Dust Aggregation with Stochastic Charging

Lorin S. Matthews[1], Babak Shotorban[2], and Truell W. Hyde[1]

[1] Center for Astrophysics, Space Physics, and Engineering Research, Baylor University, Waco, TX 76798, Lorin_Matthews@baylor.edu

[2] Department of Mechanical and Aerospace Engineering, The University of Alabama in Huntsville, Huntsville, AL 35899

Abstract

The coagulation of cosmic dust grains is a fundamental process which takes place in astrophysical environments, such as presolar nebulae and circumstellar and protoplanetary disks. Cosmic dust grains can become charged through interaction with their plasma environment or other processes, and the resultant electrostatic force between dust grains can strongly affect their coagulation rate. Since ions and electrons are collected on the surface of the dust grain at random time intervals, the electrical charge of a dust grain experiences stochastic fluctuations. In this study, a set of stochastic differential equations is developed to model these fluctuations over the surface of an irregularly-shaped aggregate. Then, employing the data produced, the influence of the charge fluctuations on the coagulation process and the physical characteristics of the aggregates formed is examined. It is shown that dust with small charges (due to the small size of the dust grains or a tenuous plasma environment) are affected most strongly.

1. INTRODUCTION

The coagulation of cosmic dust grains is a fundamental process which takes place in astrophysical environments, such as presolar nebulae and circumstellar and protoplanetary disks. Sub-millimeter observations of T-Tauri stars give evidence that as YSOs (young stellar objects and associated disk material) evolve, there is a greater population of large dust grains in the disk, which presumably grow through the agglomerations of smaller grains (Andrews and Williams 2005). The processes by which these aggregates grow to km-sized planetesimals, the precursors to planets, are still poorly understood, though much numerical and experimental work has been done to provide a physical description of the earliest stages of planetesimal formation (see e.g. Blum and Wurm 2008, Okuzumi 2009, Okuzumi *et al* 2011a, 2011b, Matthews *et al* 2012 and the references therein).

One of the key factors in the growth of aggregate grains is their internal structure. At low velocities, colliding dust particles tend to stick at their point of contact without restructuring (Dominik & Tielens 1997, Dominik et al. 2007), producing fluffy aggregates which are strongly coupled to the ambient gas. Brownian motion, vertical settling, and turbulence are mechanisms which provide a relative velocity of the grains with respect to the gas and to each other, allowing the grains to collide and grow. This growth may continue until the aggregates are up to centimeter-sizes, at which point the collision energy is enough to cause compaction (Suyama *et al* 2008). One measure of the amount of open space within an aggregate consisting of many spherical monomers is the compactness factor (Paszun and Dominik 2009)

$$\Phi_\sigma = \frac{\sum_{j=1}^{N} a_j^3}{R_\sigma^3}, \qquad (1)$$

with $N$ the number of monomers in an aggregate, each with radius $a_j$, and $R_\sigma$ the radius of the aggregate's projected cross section averaged over many orientations, referred to as the equivalent radius.

The turbulence within accretion disks is attributed to magnetorotational instability (MRI; Balbus and Hawley 1991). The MRI requires an ionized medium to couple the magnetic field to the disk material. The dust in these ionized regions become charged through the collection of electrons and ions, affecting not only the region of turbulence (Sano *et al* 2000, Okuzumi 2009) but also the interactions between the grains during the collision process and their resultant morphology (Matthews *et al* 2012, Matthews and Hyde 2008, Okuzumi 2009, Okuzumi *et al* 2011a, 2011b).

In most cases, grains charge to a negative potential due to greater electron mobility as compared to the ions. Thus, the Coulomb interaction between grains in general is repulsive, inhibiting grain growth, though mechanisms which are capable of producing positively and negatively charged grains trigger runaway growth (Ivlev *et al* 2002, Konopka *et al* 2005, Matthews and Hyde 2004). However, the coagulation rate for such like-charged grains may be enhanced if charge-dipole interactions, which can be created by anisotropic plasma flow (Lapenta 1998) or induced dipoles (Simpson 1978), are included. The physical geometry of the aggregates can also lead to a net dipole moment on the grain, which has been shown to enhance the coagulation rate for like-charged grains (Matthews and Hyde 2008, 2009).

The effect of the dust charging on aggregate growth depends on the relative densities of the gas and dust. High dust densities lead to electron depletion, and the resultant difference in grain charge has been shown to alter the physical characteristics of the dust aggregates (Matthews *et al* 2012). More highly charged dust aggregates tend to be more massive, as they preferentially incorporate the largest monomers from the initial dust population, but also tend to be fluffier, with lower compactness factors and larger equivalent radii. Numerical simulations of the coupling between the dust charge and gas ionization states have shown that porous aggregate growth eventually "freezes out" over a large portion of a protoplanetary disk, halting further dust growth, though this may be prevented by global dust transport (Okuzumi *et al* 2011a, 2011b).

Given these effects, a complete understanding of the microphysics underlying the charging and aggregation of dust is still needed. One aspect of dust charging that becomes important in weakly ionized gas environments is the stochastic nature of the charging process as ions and electrons are collected on the surface of the grain at random time intervals. This stochastic behavior is intrinsic noise (Van Kampen 2007) which occurs in systems with discrete nature. Various approaches, based on a time series approach (Cui and Goree 1994), master or Fokker-Planck equations (Matsoukas and Russell 1995, 1997, Shotorban 2011, 2012) and the Langevin equation (Khrapak *et al* 1999, Shotorban 2011), have been proposed to model the stochastic charging of spherical dust grains. Shotorban (2011) showed all of these approaches can be developed from the master equation that governs the state of the grain charge, assuming that the charging is a Markov process.

This study examines how the stochastic charge fluctuations alter the coagulation process and the physical characteristics of the aggregates formed. Section 2 describes the charging model and the manner in which the stochastic fluctuations are taken into account. Section 3 describes the model used to simulate the collisional growth of fractal aggregates and its integration with the charging model. The resultant aggregate morphology is compared for aggregation with stochastic and non-stochastic charging in Section 4, with a summary and conclusion presented in Section 5.

2. CHARGING MODEL

The charge on the surface of an aggregate can be found using orbital motion limited theory employing a line of sight approximation, OML_LOS (Matthews *et al* 2012). OML theory of cylindrical and spherical Langmuir probes is commonly used to derive the charge on a single grain particle immersed in plasma (Allen 1992). The current density due to incoming ions or electrons (plasma species *s*) to a point on the surface of a grain is given by

$$J_s = n_s q_s \int_{v_{min}}^{\infty} v_s^3 f(v_s) dv_s \iint \cos \alpha \, d\Omega \qquad (2)$$

with $n_s$ the plasma density very far from the particle, $q_s$ the charge of the incoming plasma particle, $f_s(v_s)$ the velocity distribution function of the plasma particles (assumed to be Maxwellian) and $v_s \cos(\alpha)$ the velocity component of the incoming plasma particle perpendicular to the surface. For a plasma particle having charge of the same sign as the dust grain, the minimum velocity for the particle to collide with the dust is $v_{min} = \sqrt{2q_s\Phi/m_s}$, where $\Phi$ is the electric potential on the dust grain and $m_s$ is the mass of the plasma particle. For grains with opposite charge, the minimum velocity is zero. Electrons and ions are assumed to have sufficient velocity that deviations from straight-line paths are small.

In calculating the current density to a spherical grain, plasma particles are allowed to impinge upon any given point on the grain surface from all directions except for those blocked by the grain itself; thus, the limits for angular integration are $0 \leq \theta \leq \pi/2$ and $0 \leq \phi \leq 2\pi$ providing an analytical solution for the current density. For an irregular aggregate consisting of many spherical monomers, lines of sight may be blocked by other monomers in the aggregate, and thus the integral over the angles, the LOS factor, must be determined numerically. To accomplish this, the surface of the aggregate is divided into many patches, and the LOS factor for each patch is determined by finding the open lines of sight from the center of the patch. The current density to each patch is calculated as a function of the electric potential at that patch, due to the charge on all of the patches. The charge on the aggregate is then found from the total current to the grain (the sum of the current densities to each patch multiplied by the patch areas) multiplied by the time interval $\Delta t$. This process is iterated in time until the current to the grain is essentially zero, i. e., the grain charge is changing by less than 0.01% over several time steps, to obtain the equilibrium charge.

### 2.1  Stochastic Charging

The charge collected on each patch will experience fluctuations since both ions and electrons are attached to the patch at random times. Since the ion and electron currents to each patch as a function of the charges of the patches of the aggregate is known (as described above), the charge fluctuations of the patches can be modeled as follows.

Defining the vector of elementary charges collected on the patches $\mathbf{Z} = \{Z_1, Z_2, \ldots, Z_M\} \in \mathbb{R}^M$, where $M$ is the total number of patches on the aggregate, (e.g., $Z_2$ is the number of elementary charges collected on the patch number 2) and $P(\mathbf{Z}, t)$ as the probability density function of a state at which patch number 1 has $Z_1$ charges, patch number 2 has $Z_2$ charges, etc., and assuming that $\mathbf{Z}$ undergoes a Markov process (Van Kampen 2007), a master equation can be developed as follows:

$$\frac{dP(\mathbf{Z},t)}{dt} = \sum_{p=1}^{M} I_{i,p}(\mathbf{Z} - \mathbf{e}_p) P(\mathbf{Z} - \mathbf{e}_p, t) + I_{e,p}(\mathbf{Z} + \mathbf{e}_p) P(\mathbf{Z} + \mathbf{e}_p, t) - [I_{i,p}(\mathbf{Z}) + I_{e,p}(\mathbf{Z})] P(\mathbf{Z}, t). \quad (3)$$

In this equation, $I_{i,p}(\mathbf{Z})$ and $I_{e,p}(\mathbf{Z})$ are the electron and ion currents, respectively, to patch number $p$, and $\mathbf{e}_j \in \mathbb{R}^M$ is the unit vector, e. g., $\mathbf{e}_3 = \{0,0,1,\ldots,0\}$. Equation (3) is a generalized form of the master equation given by Matsoukas and Russell (1995) and Shotorban (2011), and is formulated to determine the charge collected on the entire surface of the grain, utilizing the total ion and electron currents to the grain. As such, Equation (3) may be regarded as a gain-loss equation for the probabilities of the separate charge states of the patches, where it is assumed no charge is transferred from one patch to another.

Using a Kramers-Moyal expansion (Gardiner 2004), a Fokker-Planck equation may be obtained from the master equation (3)

$$\frac{\partial P(\mathbf{Z},t)}{\partial t} = \sum_{p=1}^{M} \frac{\partial}{\partial Z_p} [I_{e,p}(\mathbf{Z}) - I_{i,p}(\mathbf{Z})] P(\mathbf{Z}, t) + \frac{1}{2} \frac{\partial^2}{\partial Z_p^2} [I_{e,p}(\mathbf{Z}) + I_{i,p}(\mathbf{Z})] P(\mathbf{Z}, t). \quad (4)$$

It can be shown (Gardiner 2004) that this equation is statistically equivalent to the following stochastic differential equation (SDE) for the charge of each patch:

$$dZ_p(t) = [I_{i,p}(\mathbf{Z}(t)) - I_{e,p}(\mathbf{Z}(t))] t + \sqrt{I_{e,p}(\mathbf{Z}(t)) + I_{i,p}(\mathbf{Z}(t))} dW_p(t), \quad (5)$$

where $Z_p(t)$ is the time dependent charge of patch $p$, $\mathbf{Z}(t)$ is the time dependent charge of the vector of patch charges, and $W_p(t)$ is a Weiner process. Equation (5), without the last term on the right hand side, represents the change in the charge of the patch when random charge fluctuations are neglected. The SDE's can be solved numerically using the Euler-Maruyama method (Gardiner 2004):

$$Z_p^{(n+1)} = Z_p^{(n)} + [I_{i,p}(\mathbf{Z}^{(n)}) - I_{e,p}(\mathbf{Z}^{(n)})]\Delta t + \sqrt{I_{e,p}(\mathbf{Z}^{(n)}) + I_{i,p}(\mathbf{Z}^{(n)})}\sqrt{\Delta t}\xi_p, \quad (6)$$

where $\Delta t$ is the time step, $Z_p^{(n)}$ is the charge of patch number $p$ at time step $n$, and $\xi_p$ is a random number with a normal distribution.

There has been also a theoretical work on the mean and variance of the global charge collected on the surface of a spherical grain at the stationary states. For Maxwellian electron and ion distributions, the mean charge on a grain is given by (Matsoukas and Russell 1995)

$$\langle Z \rangle = C \frac{4\pi\varepsilon_0 akT_e}{e^2} \ln \frac{n_i}{n_e} \left(\frac{m_e T_e}{m_i T_i}\right)^{1/2}, \quad (7)$$

where $\langle Z \rangle$ is the average charge number on a grain of radius $a$ and $n_{e,i}$, $m_{e,i}$, and $T_{e,i}$ are the electron and ion number density, mass, and temperature. In eq. (7), $C$ is a constant on the order of unity, dependent on the mass of ions and the ratio of number densities of ions and electrons. Furthermore, Matsoukas and Russell (1995) show that for equal ion and electron temperature, the variance of the grain global charge distribution is given by

$$\sigma^2 = \frac{1 - \langle Z \rangle / \tau}{2 - \langle Z \rangle / \tau} \tau, \quad (8)$$

where $\tau = 4\pi\varepsilon_0 akT_e/e^2$.

Calculating the stochastic charging of multiple surface patches as given in Eq (6), the charging history of a single spherical grain with radius $a = 0.5$ μm is shown in Figure 1. (The plasma environment assumed is representative of conditions in a protoplanetary disk, with the electron and ion densities, $n_e$ and $n_i$, set as discussed below.) As shown, the charge fluctuates about the equilibrium charge (Figure 1a) with a normal distribution (Figure 1b). As expected, the average charge $\langle Z \rangle$ and variance are linearly proportional to the grain radius (Figure 2), though the variance in the grain charge tends to be greater than that predicted by Eq. 8, especially for the larger grain charges.

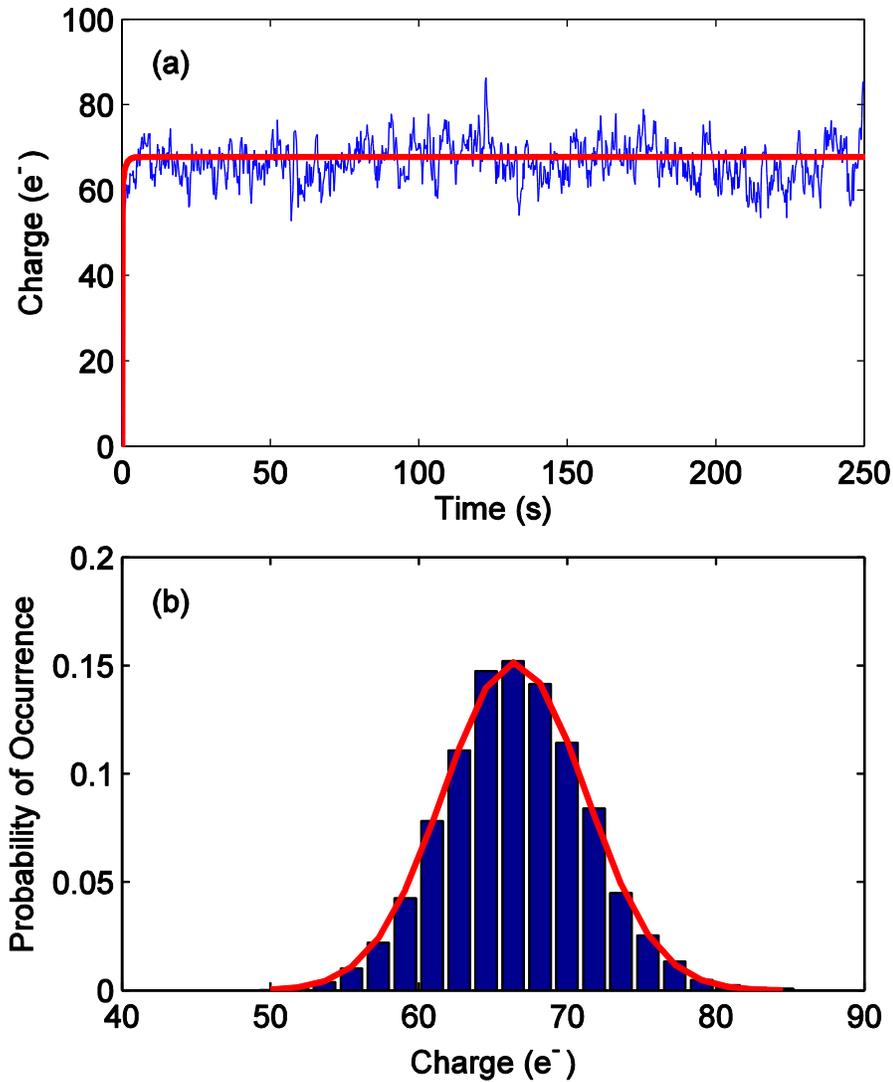

Figure 1. (a) Stochastic charging history of a spherical grain with radius $a = 0.5$ μm for a plasma with $n_e = n_i = 5 \times 10^8$ m$^{-3}$ and $T_e = T_i = 900$ K. The charging history of the same grain under the same plasma conditions without the stochastic effects taken into account is shown by the gray (red online) line. (b) Probability distribution of the charge on the grain. The histogram is obtained from the charge at more than 10,000 time steps after the grain has reached its equilibrium charge. The continuous gray line (red online) represents a normal distribution with the mean and variance obtained from simulation.

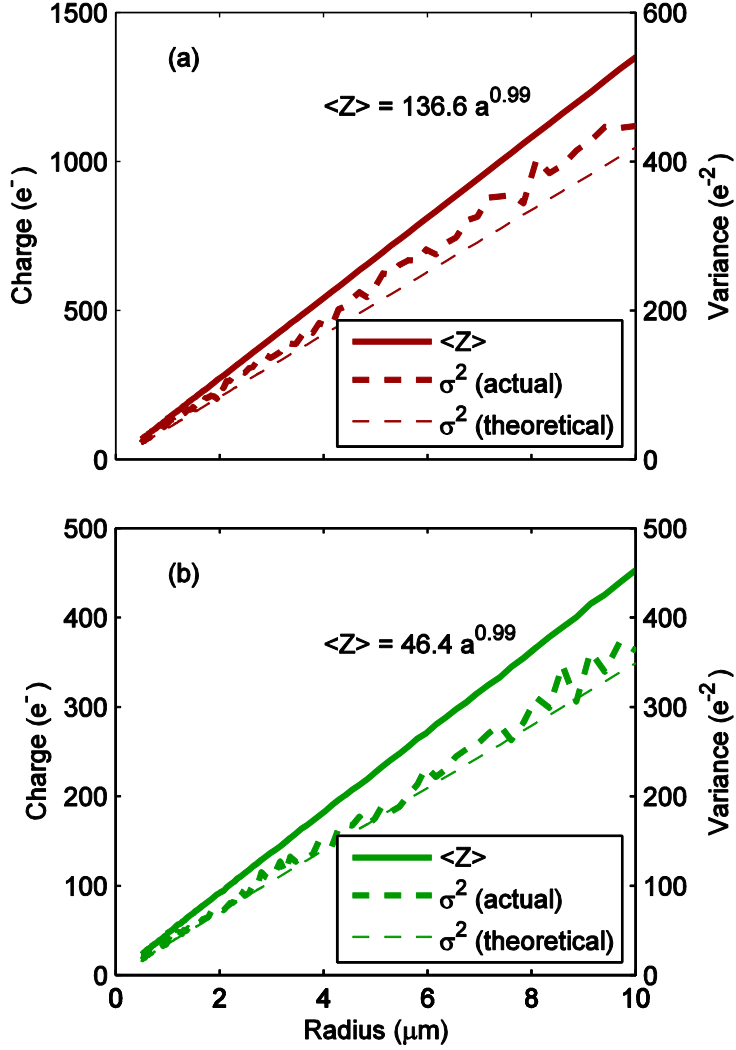

Figure 2. Mean charge collected on a spherical grain (solid line) and the variance (dashed lines) as a function of grain radius (color online). The mean and variance were calculated from 10,000 time steps after a grain had reached its equilibrium charge. The variance in the charge has a linear trend, but tends to be greater than the variance predicted by theory (Eq. 8). Plasma conditions are (a) $n_e = n_i = 5 \times 10^8$ m$^{-3}$, (b) $n_e = 0.1 n_i$.

3. AGGREGATION WITH STOCHASTIC CHARGING

Grain coagulation was modeled employing an initial population of silicate spheres having radii from $0.5 \leq a \leq 10$ μm and a power law size distribution $n(a)da \propto a^{-3.5} da$. Plasma conditions were chosen to be a simple model representative of a possible protoplanetary disk environment, with hydrogen plasma having equal electron and ion temperature, $T_e = T_i = 900$ K.

For a dusty plasma, the quasineutrality condition is $Z_e n_e + Z_i n_i + Z_d n_d = 0$, where $n_e$, $n_i$, and $n_d$ refer to the electron, ion, and dust densities, respectively, and $Z_{e,i,d}$ is the average charge number of each species. Two different plasma densities were used, one corresponding to low dust densities, so that a negligible percentage of the electrons reside on the dust grains, $n_e = n_i = 5 \times 10^8$ m$^{-3}$, and one in which the dust density is large enough that the electrons in the plasma are depleted, $n_e = 0.1\ n_i$ (Matthews *et al* 2012). In the second case, the dust charge is reduced as compared to the first (see Figure 2). The relative velocities between two interacting dust grains was set assuming the grains were small enough to be strongly coupled to turbulent eddies in a protoplanetary disk (Ormel and Cuzzi 2007). Thus, the relative velocity between grains is proportional to the difference in their stopping times, $t_s = 3m/4c_g\rho_g\sigma$, where $c_g$ and $\rho_g$ are the sound speed in the surrounding medium and the mass density of the surrounding medium, respectively, $m$ is the dust particle mass and $\sigma$ its projected cross section, here defined as $\sigma = \pi R_\sigma^2$.

As the modeled interaction time between two colliding grains is very short compared to the charging time, the charge on each grain was assumed to be constant throughout the interaction process. Using the data above, incoming monomers were given a charge based on a normal deviation from the mean charge found for stochastic charging. Upon a successful collision, the resulting aggregate was stochastically charged using OML_LOS for a set number of time steps after equilibrium was reached and saved to a library. Thus, the charge on each aggregate was randomly set to some deviation from the average charge.

Typically, growth of porous aggregates is characterized by the types of collisions modeled. In BPCA (ballistic particle-cluster agglomeration), growth proceeds through the addition of single monomers to a cluster, resulting in dense aggregates. In BCCA (ballistic cluster-cluster aggregation), collisions occur between clusters of the same size (number of monomers) (Meakin 1991). Quasi-BCCA (QBCCA) growth has also been modeled, in which collisions occur between two clusters with a fixed mass ratio (Okuzumi *et al* 2009). In this study, aggregates are created through collisions between an initial polydisperse collection of pherical monomers. As each aggregate is created, it is saved in a library and subsequent aggregates are grown through collisions between these aggregates or monomers. In this way, collisions between all possible mass ratios are sampled, with the charge and morphology of each aggregate resolved. The aggregates were built in three generations, with the first generation grown to a size $N = 20$ monomers through BPCA collisions, and the second generation grown to a size $N \sim 200$ through collisions between monomers (60% probability) or first generation aggregates (40% probability). Third generation aggregates were grown to a size $N \sim 2000$ through collisions with monomers (50%), first generation aggregates (30%) or second generation aggregates (20%).

4. AGGREGATION RESULTS

Results for the aggregates produced as described above are shown below. The data are compared for stochastic and non-stochastic charging for the two different plasma conditions (denoted as "High Dust Charge" for $n_e = n_i$ and "Low Dust Charge" for $n_e = 0.1\ n_i$) and neutral aggregates.

Figure 3 shows the charge on the aggregates as a function of their equivalent radius, $R_\sigma$, the average radius of the projected cross section of an aggregate. As expected, only a small difference can be seen in the average aggregate charge when considering charging as a stochastic process (Fig 3a). Fit lines for each of the four models (high dust charge/stochastic, high dust charge/non-stochastic, low dust charge/stochastic, and low dust charge/non-stochastic, respectively) exhibit essentially the same slope, which is a result of the LOS factor being closely related to an aggregate's physical characteristics (Ma *et al* 2013). The dipole moments on the aggregates are also similar in both charge cases (Figure 3b); however stochastic charging does lead to a difference in the slopes of the fit lines, with the non-stochastically charged dust having a steeper slope. These differences are small compared to the scatter in the data. It should be noted that the non-stochastically charged aggregates with small dust charge did not grow to as large a size; this will be discussed below.

      The variance in the charge on the aggregates is shown in Figure 4, with the variance calculated using Eq. 8 overlaid for comparison. While this line is a good fit for the average variance, the scatter in the data points is much larger than the scatter in the mean charge (Figure 3), covering more than an order of magnitude. In Figure 4a, the data points for $R_\sigma < 3$ µm have a variance considerably larger than the mean variance predicted by theory. These points mainly represent $N = 2$ aggregates, and in the highly-charged case consist of monomers which have a large difference in radii, unlike the low-dust charge case where the $N = 2$ aggregates can consist of monomers of the same radii. Thus the greater variance is due to the difference in monomer radii, and may also reflect the fact that $R_\sigma$ is not a good defining characteristic for these small aggregates.

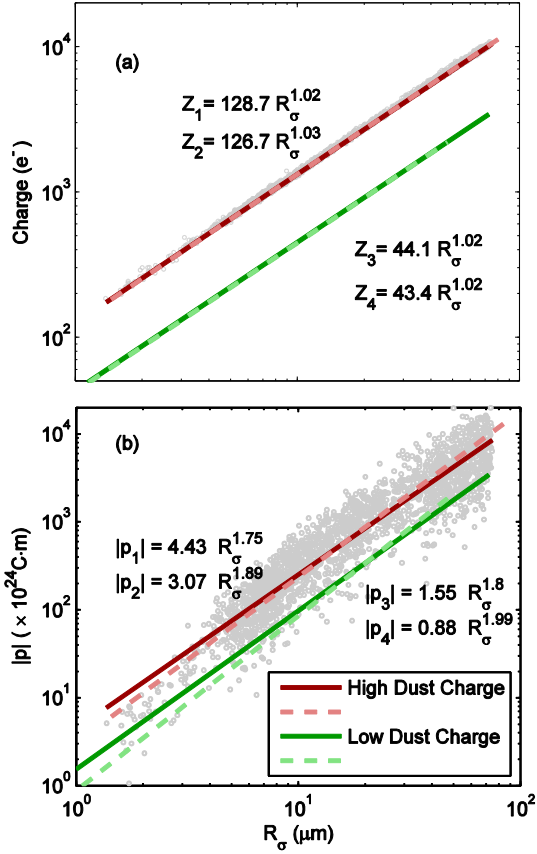
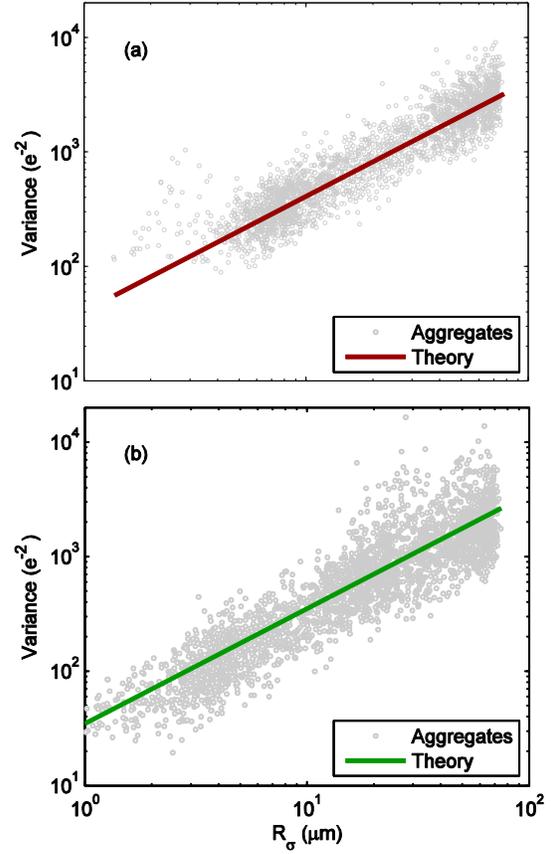

Figure 3. Aggregate charge (a) and dipole moment (b) as a function of the equivalent radius. Lines shown are linear log-log fits to the charge on all of the aggregates. The solid lines indicate the results from stochastic charging, while the lighter-color, dashed lines indicate the results of non-stochastic charging (color online). Data points for ~12,000 aggregates from the high dust charge, stochastic case are shown to indicate the spread in the data. The subscripts on $Q$ and $|p|$ refer to the charging conditions, with 1, 2 = high dust charge, stochastic/non-stochastic charging, respectively, and 3, 4 = low dust charge, stochastic/non-stochastic charging, respectively.

Figure 4. Variance in charge on aggregates for (a) $n_e = n_i$ and (b) $n_e = 0.1\, n_i$. The variance for each aggregate is calculated from the charging history after the aggregate. The line represents the variance as calculated by Eq. 8, with $<Q>$ given by the equations for the fit lines in Figure 3a.

The differences between the two charging processes become more apparent when the physical characteristics of the aggregates are examined. Figure 5 shows the monomer distribution within the third generation aggregates. The initial distribution of monomers is set as a power law with index $\alpha = -3.5$. Charged aggregates have a larger number of large monomers incorporated into the aggregates than do aggregates built from uncharged dust, with this number increasing as the grain charge increases. This is a result of the large relative velocities needed in order to overcome the coulomb repulsion barrier, with turbulent relative velocities depending on the differences in the grain radii. For the highly charged dust (plasma densities $n_e = n_i$), stochastic charging causes very little difference in this distribution. However, for the less highly charged dust (plasma densities $n_e = 0.1\ n_i$), stochastic charging flattens the monomer distribution within the aggregates, similar to the highly-charged populations.

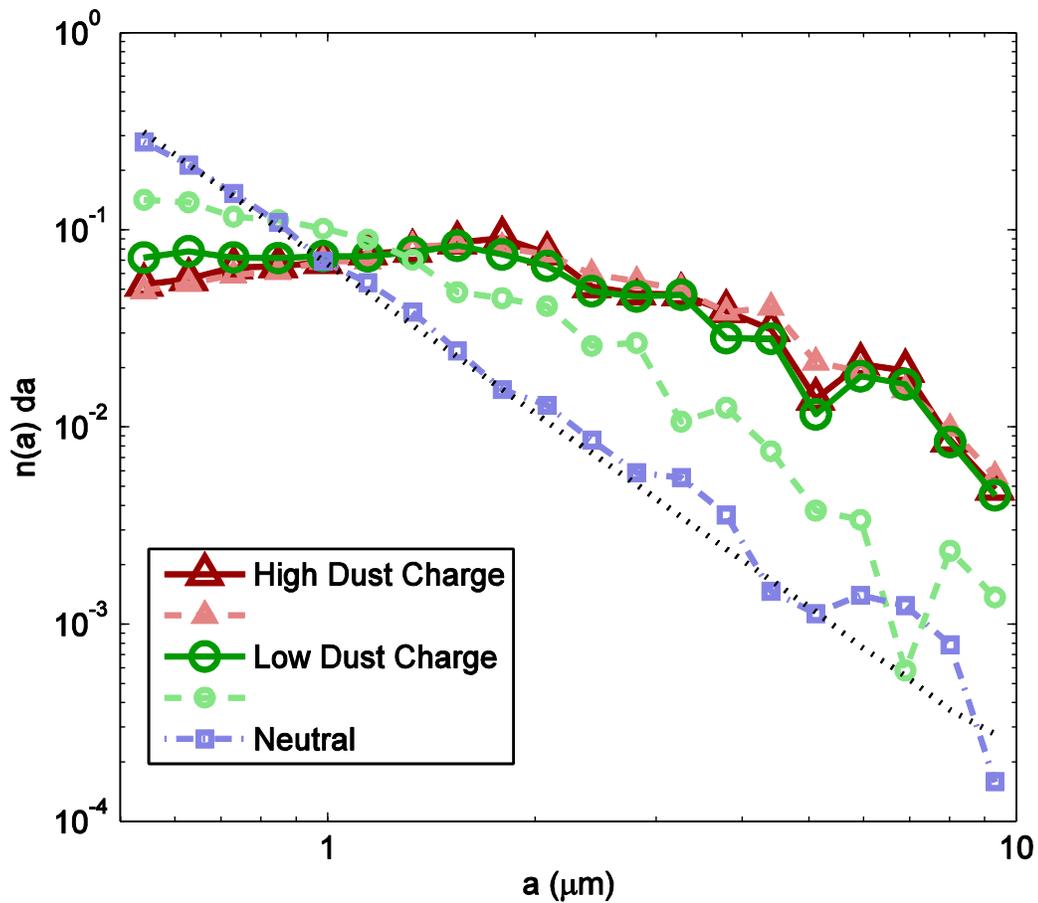

Figure 5. Size distribution of monomers within third generation aggregates (color online). The heavy, darker symbols represent stochastic charging cases while results for aggregation without stochastic effects are shown by the smaller, lighter symbols. All aggregates were built from an initial population with a power law distribution, as indicated by the dotted line. The monomers within uncharged aggregates built from this distribution closely follow this trend. Charged aggregates contain a greater proportion of large monomers. While stochastic charging has little effect for the highly charged aggregates, the opposite is true for slightly charged aggregates.

This has an effect on the physical characteristics of the aggregates, as well, as shown in Figure 6. More highly charged dust results in aggregates with greater mass (Figure 6a), a larger equivalent radius (Figure 6b), and a smaller compactness factor (Figure 6c) for a given number of monomers contained within the aggregate ($N$). The greater aggregate mass for a given number of monomers is directly related to the distribution of monomers within the aggregates. Thus, aggregates in the two populations of highly charged aggregates and the stochastically charged population of dust with low charges have similar masses. Stochastic charging effects also cause the aggregates in the low dust charge population to have larger effective radii and lower compactness factors, similar to the more highly charged aggregates. It is interesting to note that at large aggregate sizes, aggregates in both of the stochastically charged cases have very similar compactness factors, slightly more compact than the highly charged case, and slightly less compact than the low dust charge case. A probability density estimate of the compactness factor for the third generation aggregates (Figure 7) shows that the stochastically charged populations are more sharply peaked about the most probable value than the non-stochastically charged populations, as well. The physical differences are easily seen in three sample aggregates shown for the low-charge non-stochastic, high-charge stochastic, and high-charge non-stochastic cases in Figure 8. Each of the aggregates has approximately 2000 monomers and is drawn on the same scale. The low-charge non-stochastic aggregate (Fig 8a) is much smaller in size and contains many tiny monomers, resulting in the most compact configuration. The high-charge non-stochastic aggregate (Fig 8c) is much more open with a lower compactness factor, even though it has essentially the same distribution of monomer sizes as the high-charge, stochastically charged aggregate (Fig 8b), which is similar in radial extent and fluffiness to the largest aggregates in the low-charge, stochastic case.

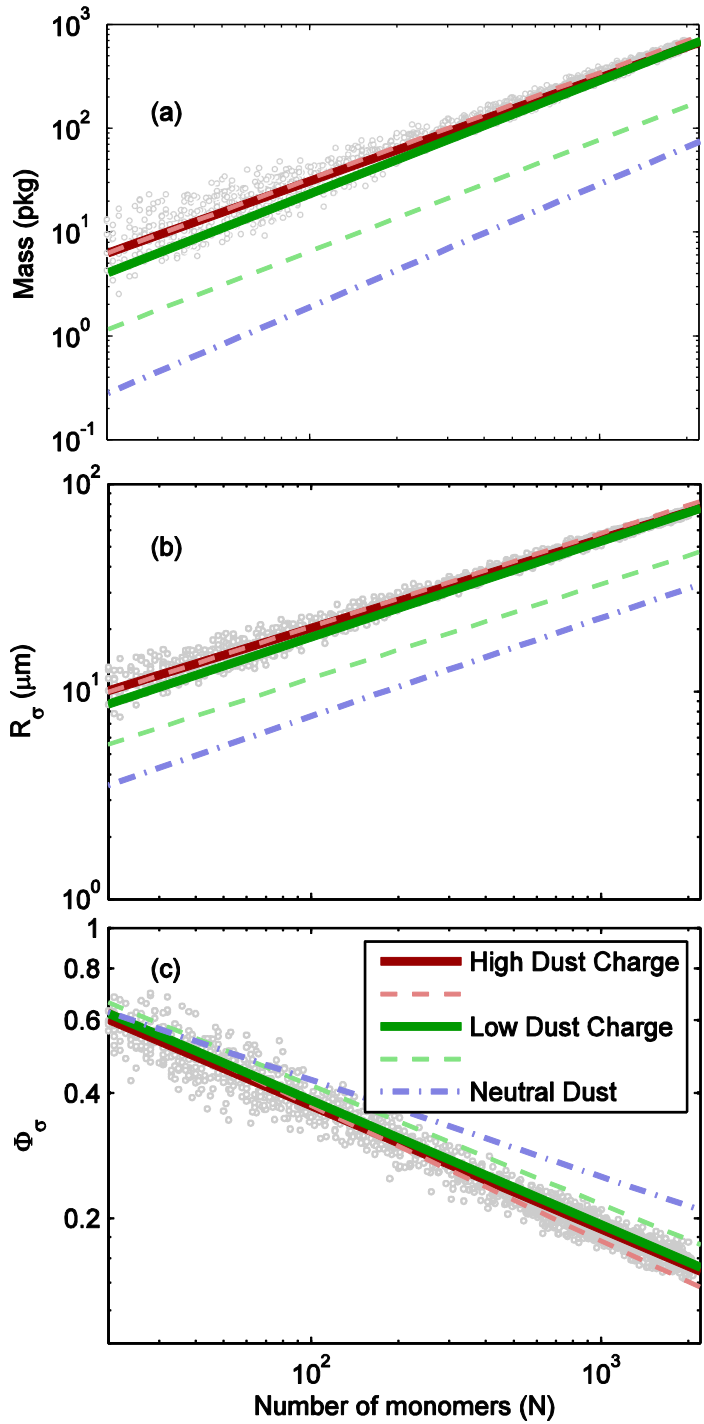

Figure 6. Mass (a), equivalent radius (b), and compactness factor (c) as a function of the number of monomers in an aggregate. Data points shown for the high dust charge/stochastic charging case. Lines represent linear log-log fits to each of the models, with the solid lines indicating stochastic charging and the lighter, dashed lines indicating the non-stochastic charging cases (color online).

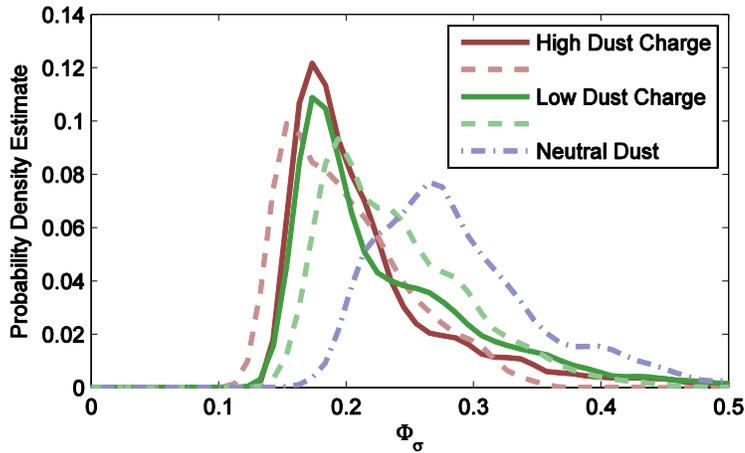

Figure 7. Probability density estimate of the compactness factor for the third generation aggregates. The solid lines indicate stochastic charging and the lighter, dashed lines indicate the non-stochastic charging cases (color online).

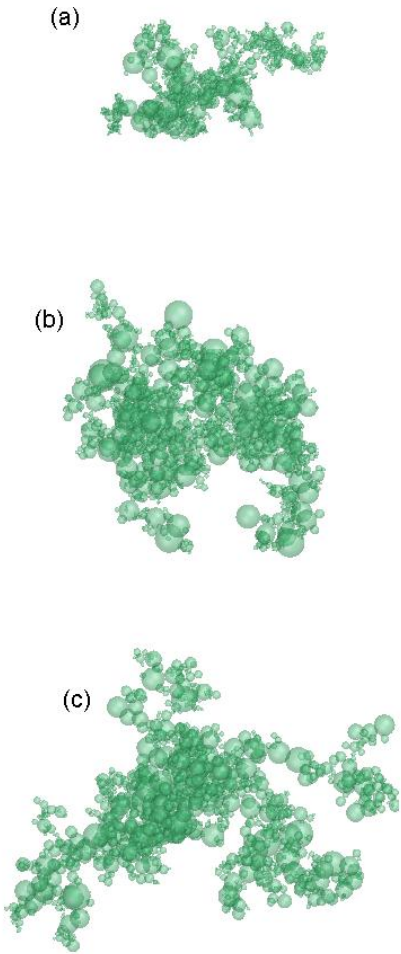

Figure 8. Representative aggregates built with a) low dust charge, non-stochastic charging, N = 2015, $R_\sigma$ = 44.4 μm, $\Phi_\sigma$ = 0.176; b) high dust charge, stochastic charging, N = 2031, $R_\sigma$ = 74.7 μm, $\Phi_\sigma$ = 0.157; and c) high dust charge, non-stochastic charging, N = 2006, $R_\sigma$ = 81.2 μm, $\Phi_\sigma$ = 0.133.

Figure 9 shows the collision probability as a function of the target aggregate size (the number of monomers within an aggregate). As shown, stochastic charging has little effect on the collision probability for the generation 1 aggregates, where aggregates grow by the addition of a single monomer. However, there is a significant difference in the collision probability for second and third generation aggregates. Stochastic charging increases the collision probability in generations 2 and 3 for the highly charged dust, but decreases the collision probability for the low dust charge case.

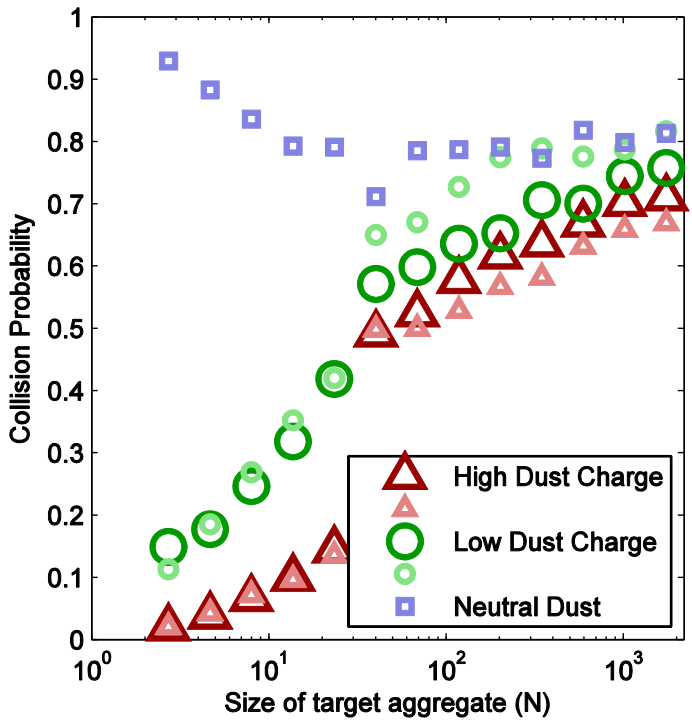

Figure 9. Collision probability as a function of number of monomers within the target aggregate. The smaller, lighter symbols indicate cases with non-stochastic charging. Collision probabilities are strongly affected by the type of collisions: first generation aggregates ($N \leq 20$) only have collisions between single spherical monomers and a target aggregate, while second ($20 < N \leq 200$) and third ($200 < N \leq 2000$) generation aggregates have collisions between both individual monomers and smaller aggregates.

Figure 10 shows the average size (number of monomers) successfully integrated into larger aggregates, as a function of the resulting aggregate size. Stochastically charged dust produced larger aggregates that could be incorporated into the second generation aggregates as compared to the corresponding non-stochastically charged cases. This appears to be due to the fact that stochastically charged dust aggregates were less likely to successfully collide with monomers (Figure 11a). The opposite trend is seen for third generation aggregates, however, with the size of an added aggregate trending towards a constant size. By examining the type of successful collisions for the third generation aggregates (Figure 11b), it can be seen that the stochastically charged dust grains were more likely to incorporate first generation aggregates, but less likely to incorporate second generation aggregates. Thus the aggregates that were swept up tend to be of a similar size regardless of the size of the target aggregate. In addition, it can be seen that for the stochastically charged dust, the highly-charged grains were more likely to incorporate monomers, while the low-charged grains less likely to incorporate monomers.

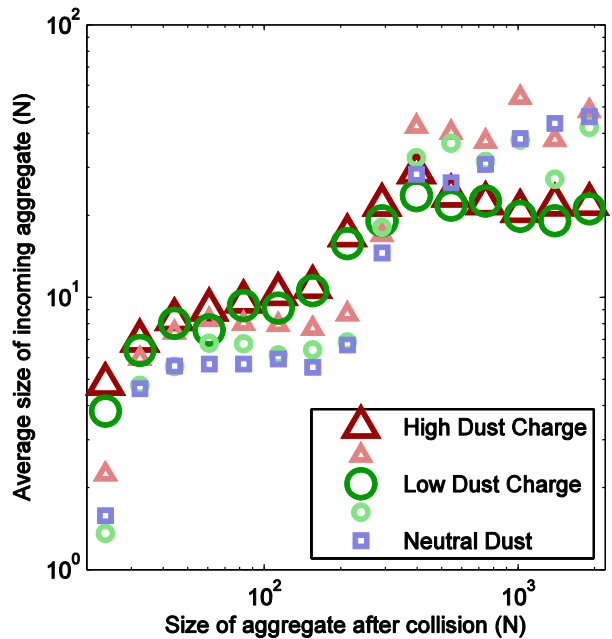
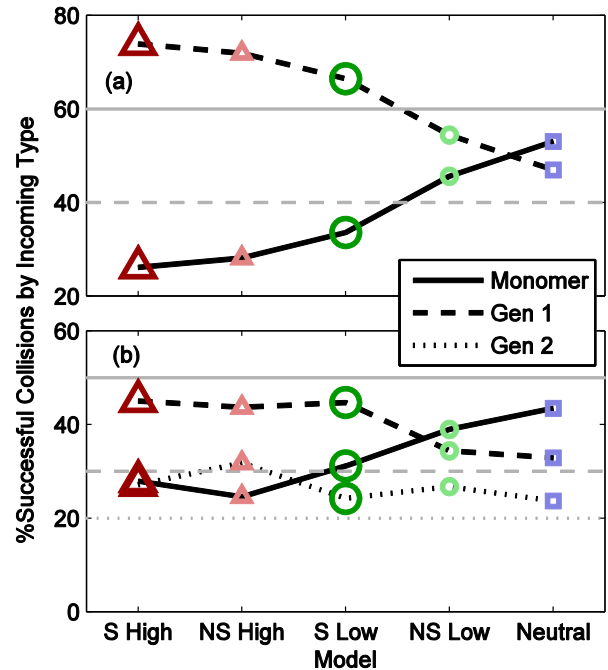

Figure 10. Average size of incoming grain successfully incorporated into an aggregate. Results are shown only for second and third generation aggregates, with the data for non-stochastic charging indicated by the smaller, lighter symbols.

Figure 11. Percentage of successful collisions (out of all successful collisions) by incoming particle type for (a) second generation aggregates and (b) third generation aggregates. Light gray lines indicate target percentages for the incoming particles: second generation aggregates are built by selecting an incoming monomer 60% of the time and an incoming aggregate 40% of the time, while third generation aggregates are built by selecting a monomer 50% of the time, a first generation aggregate 30% of the time, and a second generation aggregate 20% of the time.

## 5. CONCLUSION AND DISCUSSION

The aggregation of charged dust particles was modeled including the effects of stochastic charging fluctuations. Two different dusty plasma environments were modeled, with the dust grains in the first case having larger charges than grains in the second case. As expected, it was observed that stochastic charging effects were greatest for grains having very small charge.

In the case of more highly charged dust, stochastic charging had a very small effect on the physical characteristics of the dust aggregates formed. This behavior was expected since the charge fluctuations are less critical for dust grains with larger charge. The monomer distribution contained within aggregates (Figure 5), mass, and equivalent radius of the aggregates (Figure 6a and 6b) were shown to be identical (within the variance of the data) to those found for non-stochastic charging. In this case, stochastically charged aggregates had slightly greater compactness factors (Figure 6c). For dust with smaller charge, stochastic charging had a significant effect on the aggregation, although the average charge and dipole moment remained relatively unchanged (Figure 3). The overall characteristics of stochastically charged aggregates tended to match that of aggregates formed from highly charged dust grains, containing more of the largest monomers from the initial population (Figure 5), and exhibiting larger aggregate radii, heavier aggregate masses, and a fluffier structures for a given number of monomers (Figure 6). It is interesting to note that stochastically charged dust in both cases tended to have the same average compactness factor for the largest aggregates, and the distribution of compactness factors is more sharply peaked than that for non-stochastically charged dust (Figure 7).

Stochastic charging of grains had mixed effects on probability of collisions between grains for the largest, third generation aggregates (Figure 9). Stochastic charging increased the collision probability for the highly charged dust, but decreased the collision probability for the dust with small charges. As the stochastically charged aggregates grow to greater numbers of constituent monomers, the average size of an aggregate incorporated into a larger one also tends to approach a constant size (Figure 10), while the average size for the non-stochastically charged aggregates continues to increase. The small ($N \leq 20$) aggregates tend to more efficient at colliding with the larger grains than do monomers (Figure 11). This is likely due to the fact that the large, fluffy aggregates and small, compact aggregates form through BPCA collisions have the greatest relative velocities and thus the greatest probability of colliding. In addition, it can be seen that for the stochastically charged dust, the highly-charged grains were more likely to incorporate monomers, making them more compact (low compactness factors), while the low-charged grains less likely to incorporate monomers, making them fluffier.

In conclusion, stochastic charging will have the greatest effect in environments where dust charge is very low – typically less than a few hundred elementary charges on monomers ranging in size from 0.5 μm to 10 μm. For this case, the physical characteristics of the resultant aggregates will be similar to those for aggregates built from more highly charged dust. Although these results were obtained for a specific set of plasma conditions, the results should be correct in general in any dusty plasma environment where the dust is expected to have a fairly small charge. Thus the effects of charged grains on aggregation may extend to greater portions of a protoplanetary disk, to regions where the charge on grains was assumed to have negligible effects on aggregation.

Given that the collision probabilities strongly depend on the relative velocities and thus the compactness parameter of the aggregates, an N-body simulation is needed to see how a local dust population evolves, especially as the largest monomers are depleted from the initial population. A detailed examination of the collisions statistics will also provide parameters necessary for using a Smoluchowski method, as a statistical approach is needed to model the evolution of a full protoplanetary disk including the effect on ionization and turbulence.

ACKNOWLEDGMENTS

This work was supported by the National Science Foundation under Grant No. 0847127 and by the 2012 Junior Faculty Distinguished Research award provided by The University of Alabama in Huntsville.

REFERENCES

Allen J E 1992 Probe theory - the orbital motion approach *Phys. Scr.* **45** 497

Andrews S M and Williams J P 2005 Circumstellar Dust Disks in Taurus-Auriga: The Submillimeter Perspective *ApJ* **631** 1134

Balbus S and Hawley J 1991 A Powerful Local Shear Instability in Weakly Magnetized Disks. I. Linear Analysis *Ap. J.* **376** 214–22

Blum J and Wurm G 2008 The Growth Mechanisms of Macroscopic Bodies in Protoplanetary Disks *Annual Review of Astronomy and Astrophysics* **46** 21–56

Cui C and Goree J 1994 Fluctuations of the charge on a dust grain in a plasma *IEEE Transactions on Plasma Science* **22** 151–8

Gardiner C W 2004 *A Handbook for the Natural and Social Sciences* (Springer, New York)

Ivlev A V, Morfill G E and Konopka U 2002 Coagulation of Charged Microparticles in Neutral Gas and Charge-Induced Gel Transitions *Phys. Rev. Lett.* **89** 195502

Van Kampen N G 2007 *Stochastic Processes in Physics and Chemistry* (Elsevier, Amsterdam)

Khrapak S A, Nefedov A P, Petrov O F and Vaulina O S 1999 Dynamical properties of random charge fluctuations in a dusty plasma with different charging mechanisms *Phys. Rev. E* **59** 6017–22

Konopka U, Mokler F, Ivlev A V, Kretschmer M, Morfill G E, Thomas H M, Rothermel H, Fortov V E, Lipaev A M, Molotkov V I, Nefedov A P, Baturin Y M, Budarin Y, Ivanov A I and Roth M 2005 Charge-induced gelation of microparticles *New J. Phys.* **7** 227

Lapenta G 1998 Effect of Dipole Moments on the Coagulation of Dust Particles Immersed in Plasmas *Physica Scripta* **57** 476–80

Ma Q, Matthews L S, Land V and Hyde T W 2013 Charging of Aggregate Grains in Astrophysical Environments *ApJ* **763** 77

Matsoukas T and Russell M 1997 Fokker-Planck description of particle charging in ionized gases *Phys. Rev. E* **55** 991–4

Matsoukas T and Russell M 1995 Particle charging in low #x2010;pressure plasmas *Journal of Applied Physics* **77** 4285 –4292

Matthews L S and Hyde T W 2008 Charging and Growth of Fractal Dust Grains *IEEE Transactions on Plasma Science* **36** 310 –314

Matthews L S and Hyde T W 2009 Effect of dipole–dipole charge interactions on dust coagulation *New J. Phys.* **11** 063030

Matthews L S and Hyde T W 2004 Effects of the charge-dipole interaction on the coagulation of fractal aggregates *IEEE Transactions on Plasma Science* **32** 586 – 593

Matthews L S, Land V and Hyde T W 2012 Charging and Coagulation of Dust in Protoplanetary Plasma Environments *ApJ* **744** 8

Meakin P 1991 Fractal aggregates in geophysics *Reviews of Geophysics* **29** 317–54

Okuzumi S 2009 Electric Charging of Dust Aggregates and Its Effect on Dust Coagulation in Protoplanetary Disks *The Astrophysical Journal* **698** 1122–35

Okuzumi S, Tanaka H and Sakagami M 2009 Numerical Modeling of the Coagulation and Porosity Evolution of Dust Aggregates *The Astrophysical Journal* **707** 1247–63

Okuzumi S, Tanaka H, Takeuchi T and Sakagami M 2011a Electrostatic Barrier Against Dust Growth in Protoplanetary Disks. I. Classifying the Evolution of Size Distribution *The Astrophysical Journal* **731** 95

Okuzumi S, Tanaka H, Takeuchi T and Sakagami M 2011b Electrostatic Barrier Against Dust Growth in Protoplanetary Disks. II. Measuring the Size of the "Frozen" Zone *The Astrophysical Journal* **731** 96

Ormel C W and Cuzzi J N 2007 Closed-form expressions for particle relative velocities induced by turbulence *Astronomy and Astrophysics* **466** 413–20

Paszun D and Dominik C 2009 Collisional evolution of dust aggregates. From compaction to catastrophic destruction *Astronomy and Astrophysics* **507** 1023–40

Sano T, Miyama S M, Toyoharu U and Nakano T 2000 Magnetorotational instability in protoplanetary disks. II. Ionization state and unstable regions *The Astrophysical Journal* **543** 486–501

Shotorban B 2011 Nonstationary stochastic charge fluctuations of a dust particle in plasmas *Phys. Rev. E* **83** 066403

Shotorban B 2012 Stochastic fluctuations of dust particle charge in RF discharges *Physics of Plasmas* **19** 053702–053702–6

Simpson I C 1978 The role of induced charges in the accretion of charged dust grains *Astrophys Space Sci* **57** 381–400

Suyama T, Wada K and Tanaka H 2008 Numerical Simulation of Density Evolution of Dust Aggregates in Protoplanetary Disks. I. Head-on Collisions *ApJ* **684** 1310

Van Kampen N G 2007 *Stochastic Processes in Physics and Chemistry* (Elsevier, Amsterdam)